# Energy level structure of diamond silicon vacancy centers in an off-axis magnetic field

Shuhao Wu[1], Xinzhu Li[1], Ian Gallagher[2], Lexington Mandachi[1], Benjamin Lawrie[2,#], and Hailin Wang[1,*]

[1]Department of Physics, University of Oregon, Eugene, OR 97403, USA
[2]Materials Science and Technology Division, Oak Ridge National Laboratory, Oak Ridge, TN 37831

## Abstract

We report the development of an experimental approach that can characterize the energy level structure of an individual silicon vacancy (SiV) center in a magnetic field and extract the key parameters for the energy level structure. This approach builds upon a theoretical model that includes effects of the static strain, dynamic Jahn-Teller coupling, and magnetic field and introduces two coupling rates, one each for the ground and the excited states, to characterize the combined effects of strain and Jahn-Teller coupling. With the use of an analytical expression for the energy level structure of the SiV center under a transverse magnetic field, these two coupling rates can be directly derived from the measurement of the frequency separation between two spin-conserved transitions in a photoluminescence excitation spectrum and the measurement of the coherent population trapping resonance related to the SiV ground spin states. Additional experimental studies on the dependence of the energy level structure on the magnetic field amplitude and direction further verify the theoretical model and reveal contributions from unequal orbital magnetic coupling for the ground and excited states.

#lawriebj@ornl.gov
*hailin@uoregon.edu

# 1) Introduction

Negatively charged silicon vacancy (SiV) centers in diamond have recently emerged as a promising qubit platform for optical quantum networks and quantum spin mechanics because of their superior optical properties, long spin coherence time at low temperatures, and strain coupling to the orbital degrees of freedom in the ground states[1-6]. The zero-phonon line of SiV centers contains about 70% of the total fluorescence, with nearly lifetime-limited optical linewidth and with the transition robust against surface charge fluctuations. Excellent optical coherence has been observed for SiV centers in diamond nanostructures and in diamond membranes as thin as 100 nm[7-11]. Decoherence time exceeding 10 ms, limited by the nuclear spin bath, has been observed at temperatures near 100 mK[5]. SiV centers have been incorporated into photonic crystal structures as well as whispering gallery mode (WGM) optical resonators for cavity QED studies and for the development of quantum networks [11-16]. In addition, the special energy level structure of SiV centers can enable relatively strong ground-state spin-mechanical interactions[6, 17], providing a promising platform for quantum spin-mechanics and the development of mechanical quantum networks[18-20]. Recent advances in this area include the realization of GHz diamond spin-mechanical resonators with mechanical Q-factors exceeding $10^6$ and with efficient orbital coupling to SiV centers[21, 22].

Microwave and acoustic quantum controls of the spin states of individual SiV centers have been well established[23, 24]. All optical control of SiV spin states have also been demonstrated with coherent population trapping or coherent Raman transitions[25-27]. In spite of these experimental advances, a persistent experimental issue is the large and seemly random variations in the microwave or acoustic transition frequencies or simply magnetic-field induced splitting of the SiV ground spin states, as evidenced in the earlier studies. This is in contrast to diamond nitrogen vacancy (NV) centers, for which the energy levels of the ground states are well characterized and controlled[28].

A SiV center is an interstitial point defect consisting of a silicon atom positioned midway between two adjacent vacancies in the diamond lattice. The energy level structure of a SiV depends on the local strain environment, including static strain and the dynamic Jahn-Teller (JT) coupling[29]. Since the ground states of a SiV center are doubly degenerate, an external magnetic field is needed to induce an energy separation between two relevant spin states. For acoustic or optical spin control, a mixing of the two spin states is also necessary, which can be induced by an off-axis magnetic field, i.e. a field away from the SiV axis. For quantum control of SiV spin states, the ground state energy level structure of the SiV center thus depends sensitively on the interplay between static strain, dynamic JT effects, and external magnetic fields, which can lead to large variations in the magnetic-field induced ground state energy level separations. While theoretical analyses have separately included JT effects, static strain, and magnetic fields, all three effects have not been included simultaneously. Earlier studies of energy level structures of low strain samples included the JT coupling and off-axis magnetic field[29], while a later study focusing on changes in SiV energy level structures induced by relatively high strain included strain and off-axis magnetic field but ignored the JT effects [6]. Given the large and inevitable variations in the magnetic-field induced ground-state energy level separations, an experimental approach for a thorough characterization and a clear understanding of the energy level structures of individual SiV centers in an off-axis magnetic field are important for SiV-based studies, such as spin-mechanics or optical spin control, and for developing experimental approaches to mitigating the variations in the ground-state energy level separations.

In this paper, we report the development of an experimental approach that characterizes the energy level structure of individual SiV centers and especially extracts the key parameters for describing the energy level structure. Our approach is based on a theoretical analysis that includes the static strain, dynamic JT coupling, and magnetic field and on the introduction of two strain/JT coupling rates (one each for the ground and the excited states) that characterize the combined effects of static strain and dynamic JT coupling. We show that the strain/JT coupling rates can be directly derived from the energy level structure of the SiV center in a transverse magnetic field, specifically from the analytical expression for the energy levels. Additional experimental studies on the dependence of the energy level structure on the magnetic field amplitude and direction further verify the theoretical model and reveal contributions from unequal orbital magnetic coupling for the ground and excited states.

## 2) Theoretical model

For a negatively charged SiV center, both the ground and excited states are characterized by orbital states, $|e_x\rangle$ and $|e_y\rangle$, and spin states, $|\uparrow\rangle$ and $|\downarrow\rangle$, with the spin-orbit (SO) coupling described by ($\hbar=1$)[29]

$$H_{so} = -\lambda_{so} L_z S_z, \tag{1}$$

where $x, y, z$ are the internal basis of the SiV, with the $z$-axis along the SiV axis, and $\lambda_{so}$ is the spin-orbit coupling rate. In the Cartesian coordinates, $X, Y, Z$, of the cubic unit cell, we can define the unit vectors of the SiV basis as $\hat{x}=[1,1,-2]/\sqrt{6}$, $\hat{y}=[1,-1,0]/\sqrt{2}$, and $\hat{z}=[1,1,1]/\sqrt{3}$. In the absence of external magnetic field and static strain and in the limit that the dynamic JT effect can be ignored, both ground and excited states of a SiV are characterized by doubly degenerate doublets, with the two upper states being $|e_+,\uparrow\rangle$ and $|e_-,\downarrow\rangle$ and the two lower states being $|e_-,\uparrow\rangle$ and $|e_+,\downarrow\rangle$, where $|e_\pm\rangle = (|e_x\rangle \pm i|e_y\rangle)/\sqrt{2}$, as illustrated in Fig. 1a. The splitting between the upper and lower states in the doublet is $\lambda_{so}$, with $\lambda_{so}^g$ and $\lambda_{so}^e$ for the ground and excited states, respectively. Optical transitions taking place between the ground- and the excited-state doublets are labeled as A, B, C, D transitions in Fig. 1a.

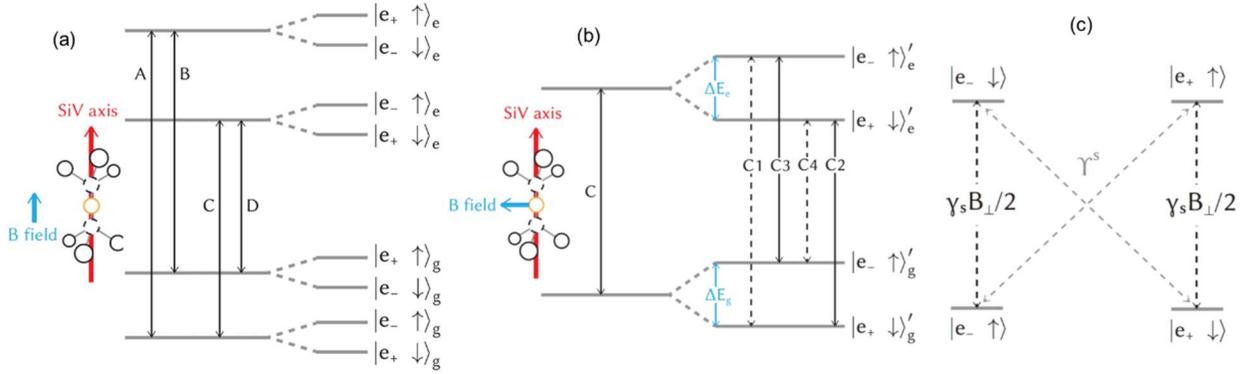

**Fig. 1.** (a) The SiV energy level structure and optical transitions in the absence of strain and JT coupling. A magnetic field along the SiV axis removes the spin degeneracy, with the optical transitions occurring between the same spin states. (b) Schematic of the energy levels and selection rules for the optical C-transition, including spin-conserved transitions (solid lines) and spin-flip transitions (dashed lines), under a magnetic field normal to the SiV axis. (c) Schematic illustrating that the transverse magnetic field induces spin-state mixing (dashed black lines), and the strain/JT coupling induces orbital-state mixing (dashed grey lines).

The interaction Hamiltonian due to an external magnetic field is given by
$$H_B = f\gamma_L B_z L_z + \gamma_S \mathbf{B}\cdot\mathbf{S}, \tag{2}$$

where $\gamma_L$ and $\gamma_s$ are the orbital and spin gyromagnetic ratio, respectively, and $f$ is the orbital quenching factor. Because of the small orbital quenching factor (~0.1), effects of the magnetic field on the orbital states are typically ignored. A magnetic field along the z axis removes the spin degeneracy of all the relevant states, with the optical transitions taking place between states with the same electron spin, as illustrated in Fig. 1a. In comparison, a transverse magnetic field (i.e., normal to the z-axis) can induce considerable spin-state mixing and thus enable spin-flip transitions (see Fig. 1b), as will be discussed in more detail later.

Strain couples only to the orbital states. With symmetry-adjusted linear combination of strain tensor components, $\alpha$, $\beta$, and $\gamma$, the strain Hamiltonian is given by (in the $|e_x\rangle$ and $|e_y\rangle$ basis)[6]

$$H_{Strain} = \begin{pmatrix} \alpha-\beta & \gamma \\ \gamma & \alpha+\beta \end{pmatrix} \otimes \begin{pmatrix} 1 & 0 \\ 0 & 1 \end{pmatrix}, \tag{2}$$

where $\alpha$ induces an overall energy shift. The JT interaction Hamiltonian is given by in the same basis as

$$H_{JT} = \begin{pmatrix} \Upsilon_x & \Upsilon_y \\ \Upsilon_y & -\Upsilon_x \end{pmatrix} \otimes \begin{pmatrix} 1 & 0 \\ 0 & 1 \end{pmatrix}, \tag{3}$$

where $\Upsilon_x$ and $\Upsilon_y$ are the JT coupling rates along $x$ and $y$, respectively. Since strain and dynamic JT coupling have the same symmetry property, it is difficult to experimentally distinguish between these two effects. We thus introduce a new set of parameters to describe the combined effects of strain and dynamic JT coupling, with $\Upsilon_x^s = -\beta + \Upsilon_x$ and $\Upsilon_y^s = \gamma + \Upsilon_y$. In the basis of the SO eigen states $\{|e_+,\uparrow\rangle, |e_-,\downarrow\rangle, |e_-,\uparrow\rangle, |e_+,\downarrow\rangle\}$, the total Hamiltonian with $f=0$ is then given by

$$H_{Total} = \frac{\lambda_{so}}{2}\begin{pmatrix} -1 & 0 & 0 & 0 \\ 0 & -1 & 0 & 0 \\ 0 & 0 & 1 & 0 \\ 0 & 0 & 0 & 1 \end{pmatrix} + \frac{\gamma_s}{2}\begin{pmatrix} B_z & 0 & 0 & B_- \\ 0 & -B_z & B_+ & 0 \\ 0 & B_- & B_z & 0 \\ B_+ & 0 & 0 & -B_z \end{pmatrix} + \begin{pmatrix} \alpha & 0 & \Upsilon_-^s & 0 \\ 0 & \alpha & 0 & \Upsilon_+^s \\ \Upsilon_+^s & 0 & \alpha & 0 \\ 0 & \Upsilon_-^s & 0 & \alpha \end{pmatrix} \tag{5}$$

where $B_\pm = (B_x \pm iB_y)$ and $\Upsilon_\pm^s = (\Upsilon_x^s \pm \Upsilon_y^s)$. As shown by the Hamiltonian, $B_x$ and $B_y$ mix spin states with the same orbital component, with the mixing ratio scaling with $\gamma_s B_\perp / \lambda_{so}$, while $\Upsilon_x^s$ and $\Upsilon_y^s$ mix orbital states with the same spin component, as illustrated in Fig. 1c.

For a transverse magnetic field, i.e., for $B_z=0$, we can analytically diagonalize the above Hamiltonian. The eigen energy of the two lower and upper states are respectively given by

$$E_{\pm}^L = \alpha - \sqrt{(\lambda_{so}/2)^2 + (\gamma_s B_{\perp}/2 \pm \Upsilon^s)^2}, \tag{6a}$$

$$E_{\pm}^U = \alpha + \sqrt{(\lambda_{so}/2)^2 + (\gamma_s B_{\perp}/2 \pm \Upsilon^s)^2}, \tag{6b}$$

where $B_{\perp} = \sqrt{B_x^2 + B_y^2}$ and $\Upsilon^s = \sqrt{(\Upsilon_x^s)^2 + (\Upsilon_y^s)^2}$, which can be viewed as the effective strain/JT coupling rate. The solution given in Eq. 6 is valid even if $f \neq 0$, since the orbital Zeeman effect is zero when $B_z=0$. Note that because of the symmetry of the SiV center, the energy level structure depends on $\Upsilon^s$ instead of $\Upsilon_x^s$ and $\Upsilon_y^s$. This is also true for a magnetic field in an arbitrary direction.

For the two lower (or upper) eigen states, effects of the transverse magnetic coupling and the strain/JT coupling cancel each other for one state and add together for the other state, as shown by Eq. 6. This asymmetry lifts the degeneracy of the two states, with the splitting (for both the lower and upper states) given by

$$\Delta E \approx 2\gamma_s B_{\perp} \cdot \Upsilon^s / \lambda_{so}, \tag{7}$$

where we assume $\lambda_{so} >> (\gamma_s B_{\perp}, \Upsilon^s)$. This splitting provides a direct measurement of the effective strain/JT coupling rate, $\Upsilon^s$. In the limit that the magnetic field is zero, the doublet remains doubly degenerate. In this case, the strain/JT coupling modifies the splitting between the upper and lower states in the doublet according to Eq. 6. Note that the above theoretical analysis applies to both the ground and excited states of the SiV center.

In the presence of a transverse magnetic field, $s_z$ is no longer a good quantum number. Spin-state mixing induced by the transverse magnetic field enables spin-flip transitions. Figure 1b illustrates the energy level structure and optical selection rules for the C transition. With $\gamma_s B_{\perp}/\lambda_{so} << 1$, spin-flip transitions (dashed lines in Fig. 1b) are expected to be much weaker than the spin-conserved transitions (solid lines in Fig. 1b). As shown in Fig. 1b, we append a "prime" to the notations for the SO eigen states to denote the new eigen states, which include effects of the spin-state mixing as well as distortions of the orbital states induced by strain/JT coupling. The magnetic field induced splitting for the ground and excited states are

$$\Delta E_g \approx 2\gamma_s B_\perp \cdot \Upsilon_g^s / \lambda_{so}^g, \tag{8a}$$

$$\Delta E_e \approx 2\gamma_s B_\perp \cdot \Upsilon_e^s / \lambda_{so}^e, \tag{8b}$$

respectively, with $\Upsilon_g^s$ and $\Upsilon_e^s$ being the corresponding effective strain/JT coupling rates. The above theoretical analysis suggests a straightforward experimental approach to determining $\Upsilon_g^s$ and $\Upsilon_e^s$, two key parameters in describing the energy level structure of a SiV. Specifically, $\Delta E_g$ and thus $\Upsilon_g^s$ can be determined directly in an experiment. The splitting between the two spin-conserved transitions, given by $\delta E = \Delta E_g - \Delta E_e$, can then be used to derive $\Delta E_e$ and thus $\Upsilon_e^s$.

### 3) Experimental results

For our experimental studies, we will use coherent population trapping (CPT) of the two lower ground states and the frequency splitting between the two spin-conserved transitions (i.e., $C_2$ and $C_3$ in Fig. 1b) in the C-transition of the SiV PLE spectrum, when the magnetic field is normal to the SiV axis, to determine the strain/JT coupling rates for the ground and the excited states. In addition, we will also use dependences of the spin-conserved splitting on the magnetic field amplitude along the Z-axis and on the orientation of the magnetic field in the X-Y plane as well as the z-Z plane to verify the theoretical model and to identify contributions from non-zero quenching factors, especially effects of unequal quenching factors of ground and excited states.

SiV centers used in our experiments were implanted approximately 75 nm below the surface of an electronic grade diamond grown by chemical vapor deposition (CVD). The average kinetic energy and dosage of the $^{28}$Si used in the implantation are 100 keV and $3\times10^9$/cm$^2$, respectively. Stepwise thermal annealing up to a temperature of 1200 °C followed by wet chemical oxidation was used for the removal of the damaged surface layer as well as for the formation of SiV centers. The sample was mounted on the mixing chamber of a dilution refrigerator (Leiden Cryogenics CFCS81-1000M), which is also equipped with a 3D vector magnet (American Magnetics). The SiV fluorescence was collected with an objective with NA=0.82 integrated into an 8f imaging system inside the refrigerator[30]. SiV fluorescence exiting an optical window at the top of the refrigerator was coupled into a multi-mode fiber with a diameter of 10 μm and then sent to an avalanche photodiode for photon counting. For photoluminescence excitation (PLE) spectra, a 532 nm laser pulse was used for the initialization of the SiV center. A red laser with a wavelength near 737 nm (New Focus Velocity TLB-6700) was used for the resonant excitation of the SiV center. An electrooptic modulator (EOM) was used to generate the two optical fields needed for the CPT experiments. One field came from the carrier wave and the other came from the first sideband generated by the EOM. Optical pulses needed were generated with acousto-optic modulators (AOMs). Experimental studies presented in this paper were performed at a temperature near 4 K and with two SiVs centers (SiV1 and SiV2) from two separate samples.

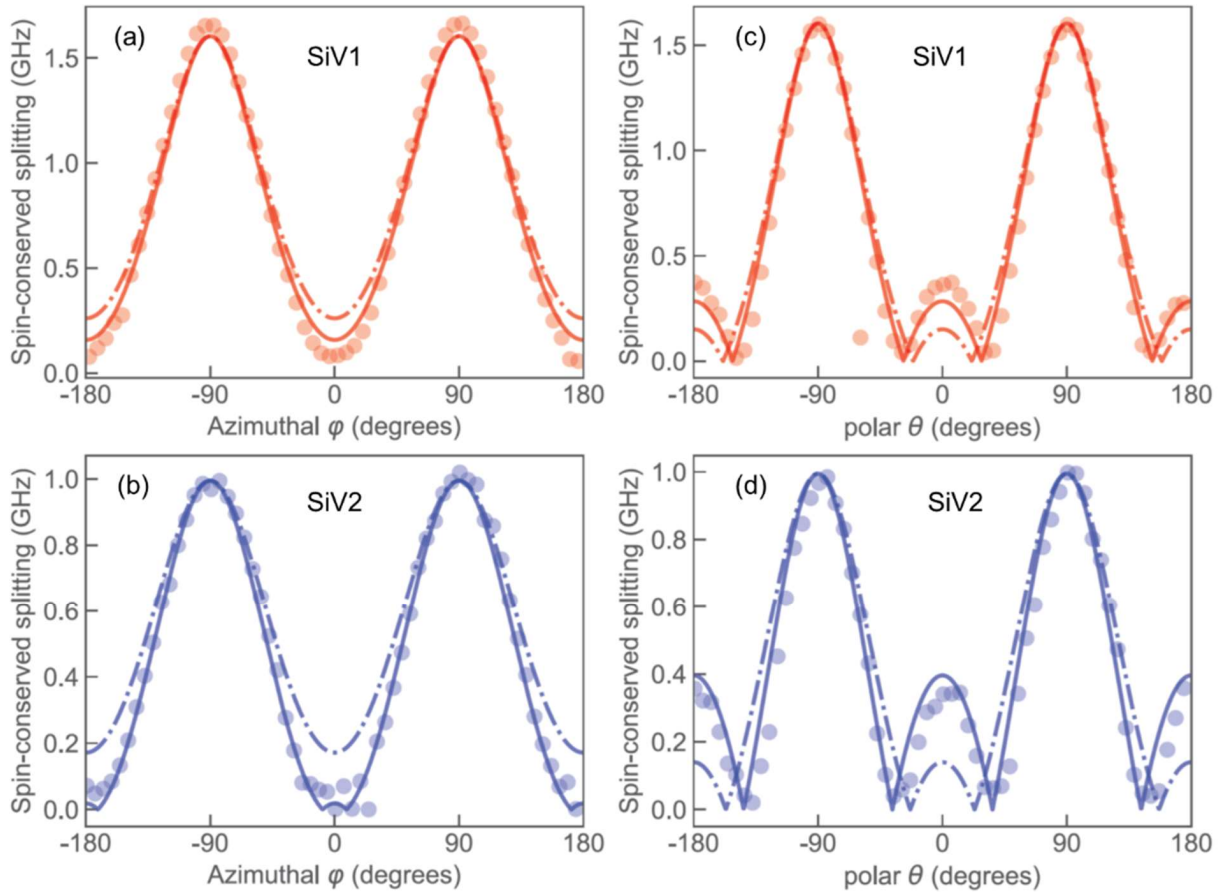

**Fig. 2.** (a) and (b) Spin-conserved splitting obtained as a function of the azimuthal angle of the B-field in the X-Y plane. (c) and (d) Spin-conserved splitting obtained as a function of the polar angle (with respect to the z-axis) of the-B field in the z-Z plane. The dotted and solid lines represent calculated spin-conserved splitting with equal and unequal orbital quenching factors, respectively, as discussed in the text.

To determine the direction of the SiV axis and especially an axis normal to the SiV axis, we measured the spin-conserved splitting derived from corresponding PLE spectra as a function of the orientation of the magnetic field in the X-Y plane with magnetic field, $B$=0.185 T, as shown in Figs. 2a and 2b. As the magnetic field rotates in the X-Y plane, the angle between the SiV axis and the field varies from 35.3 degrees to 90 degrees. Theoretically, we expect that the maximum spin-conserved splitting occurs when the magnetic field is normal to the SiV axis. Note that for convenience, we set one of the angular positions, where the spin-conserved splitting reaches the maximal value, as $\varphi$=90°. From the angular coordinates for the maximum spin-conserved splitting, we can then determine the direction of the SiV axis. We further confirmed the direction of the SiV axis by showing that the spin-conserved splitting obtained remains nearly unchanged when we rotate the direction of a magnetic field in the plane normal to the SiV axis. In addition, we also measured the spin-conserved splitting as a function of the orientation of a magnetic field in the z-Z plane with $B$=0.185 T, as shown in Figs. 2c and 2d. Note that z is along the SiV axis as discussed earlier.

Figures 3a and 3b show the spin-conserved splitting of the C-transition vs the amplitude of the magnetic field when the field is along the Z-axis for SiV1 and SiV 2, respectively. Figures 3c and 3d show the spin-conserved splitting as a function of the amplitude of the magnetic field when the field is normal to the SiV axis. Figure 4a shows, as an example, the PLE spectrum of SiV1 obtained when the magnetic field is normal to the SiV axis, with $B$=0.13 T. We will use the spin-conserved splitting shown in Fig. 4a, along with the CPT resonance position shown in Fig. 4b, to determine the strain/JT parameter for the ground and excited states.

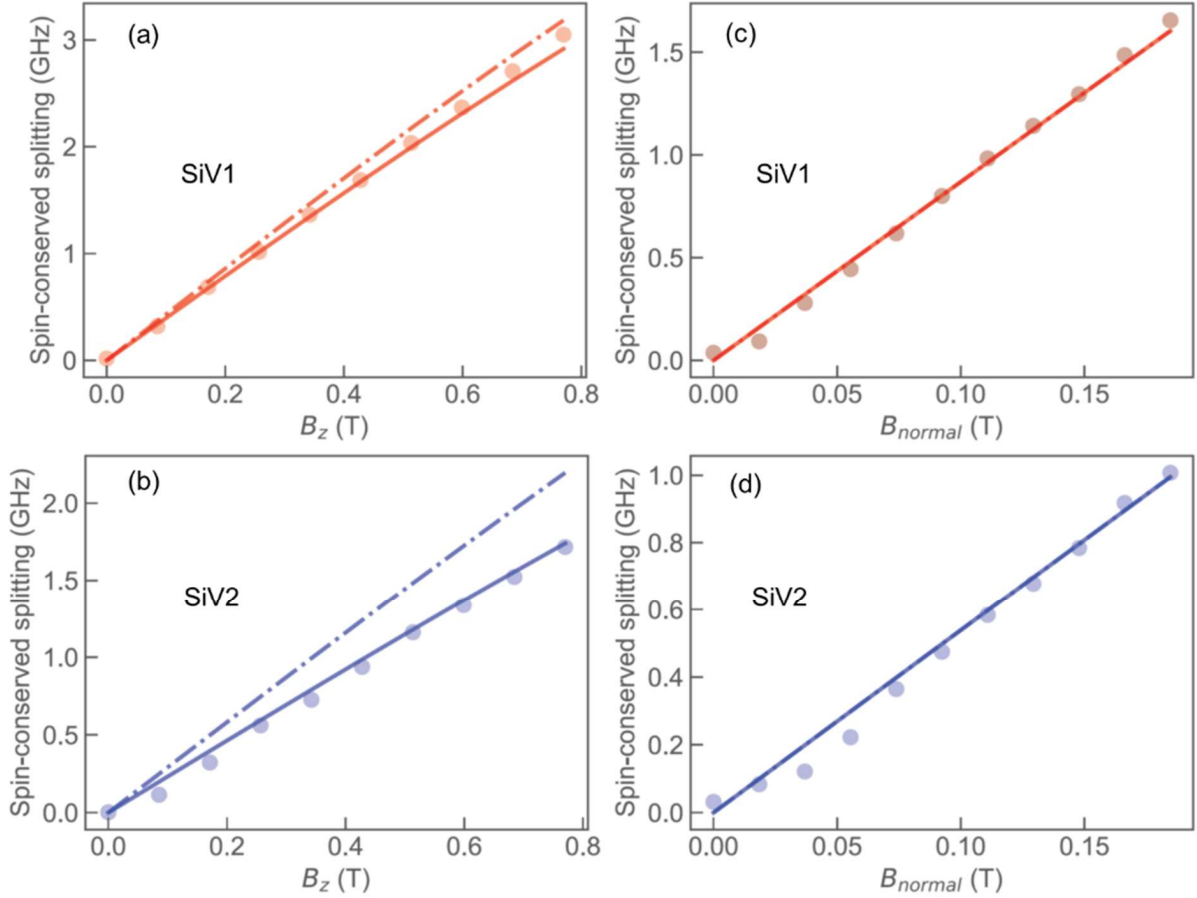

**Fig. 3.** (a) and (b) Spin-conserved splitting obtained as a function of the B-field strength and with the B-field normal to the SiV axis. (c) and (d) Spin-conserved splitting obtained as a function of the B-field strength and with the B-field along the Z-axis. The dotted and solid lines represent calculated spin-conserved splittings with equal and unequal orbital quenching factors, respectively, as discussed in the text. The two lines are identical in (c) and (d) because there is no orbital contribution when the field is normal to the SiV axis.

For the CPT experiment, we used the spin-conserved $C_2$ transition and the spin-flip $C_4$ transition in Fig. 1b to form a Λ-type three-level system. The spectral position of the CPT resonance directly measures the frequency separation of the two lower energy states in the ground state doublet. In principle, this frequency separation can also be deduced from the corresponding frequency separation between the underlying spin-conserved and spin-flip transitions in the PLE spectrum. Nevertheless, the weak spin-flip transition makes it difficult to accurately determine or even conclusively identify the transition in the PLE spectrum. To search for the CPT resonance, we first fix the optical carrier wave at the spin conserved $C_2$ transition, while the frequency of the first sideband generated by the EOM is scanned over the expected spectral range of the spin-flip transition. After the successful observation of the CPT resonance, we change the experimental setting such that the optical carrier wave is fixed at the weak spin-flip $C_4$ transition and the first sideband is now near the spin-conserved transition. This is because strong CPT dips occur when the Rabi frequencies for the spin-conserved and spin-flip optical transitions are comparable. Figure 4b shows a CPT spectral response observed for SiV1 in this setting with a magnetic field, $B$=0.13 T, normal to the SiV axis.

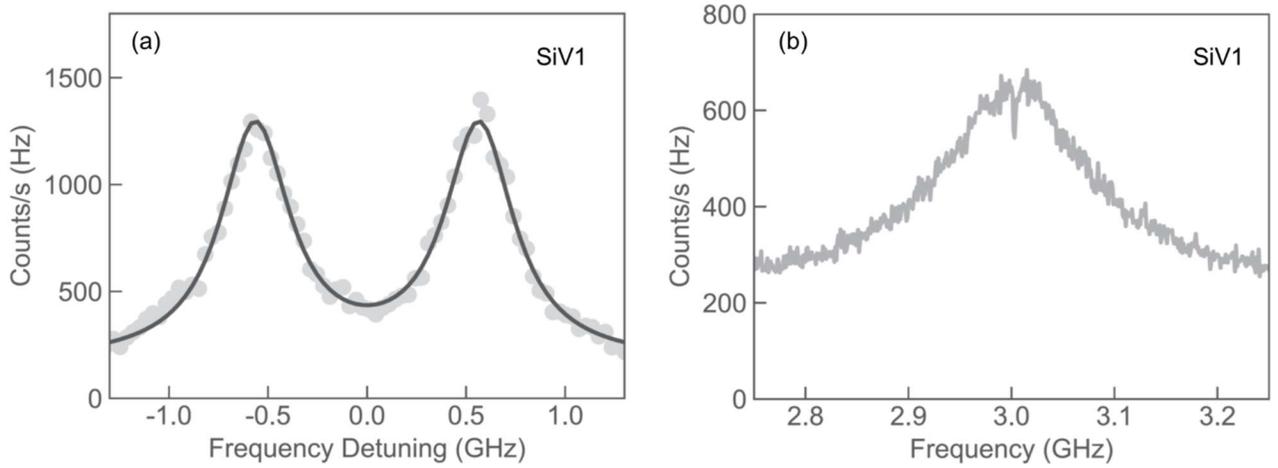

**Fig. 4.** (a) PLE spectrum of the SiV C-transition. The two peaks correspond to the $C_2$ and $C_3$ transitions. The solid line is a least-square fit to two Lorentzians. (b) CPT spectral response obtained with an input laser power of 0.25 μW. All were obtained with $B$=0.13 T along a direction normal to the SiV axis.

**4) Analysis and discussions**

The spectral position of the CPT resonance and the spin-conserved splitting such as those observed in Fig. 4, along with the use of Eq. 6a and Eq. 6b, yield $\Upsilon_g^s$ = 33, 46 GHz and $\Upsilon_e^s$ = 77, 129 GHz, for SiV1 and SiV2, respectively, where we used estimated $\lambda_{so}^g$ = 45 GHz and $\lambda_{so}^e$ = 257 GHz[29]. In comparison, earlier studies using low strain samples have obtained a maximum JT coupling rate of 15 GHz and 76 GHz for the ground and excited states, respectively, based on numerical fittings of a relatively large set of experimental results [29]. Note that since effects of strain and JT coupling cannot be separated in the PLE experiments, the effects of strain should already be included in the earlier study. The difference in the parameters obtained indicates much greater strain contributions for the SiV centers used in our study.

We have used the above strain/JT coupling rates to numerically calculate the spin-conserved splitting under the various magnetic field configurations used in Fig. 2 and Fig. 3, with the assumption of equal orbital quenching factor of 0.1 for both the ground and excited states, which has been used in all earlier studies. Note that there is no other fitting parameter used in these calculations. While there are overall good agreements, there are still noticeable and systematic deviations between the experimental and the calculated values. In particular, the calculations underestimate the spin-conserved splitting when the magnetic field is along the SiV axis and overestimate the spin-conserved splitting when the magnetic field is along the Z-axis, as shown in Fig. 2 and 3, respectively. As indicated by the solid lines in Fig. 2 and Fig. 3, both of these discrepancies can be resolved if we assume unequal quenching factors for the ground and excited states. Specifically, a quenching factor of 0.13 and 0.17 for the excited states were used for SiV1 and SiV2, respectively.

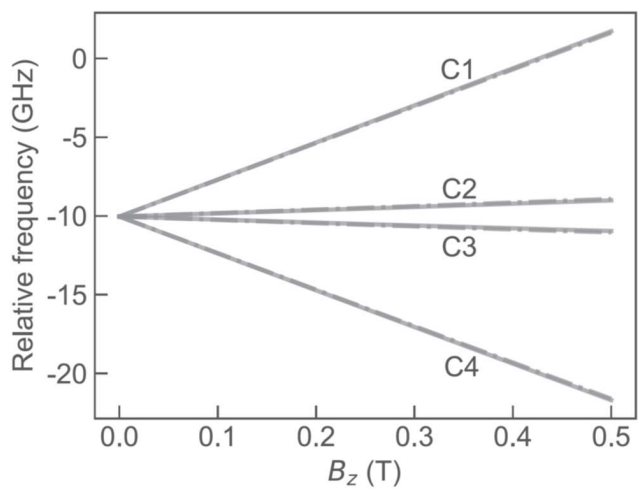

**Fig. 5.** Calculated transition frequency for the four C-transitions under a B-field along the Z axis and with the parameters given in the main text. An equal quenching factor of 0.1 was used for the dotted lines. A quenching factor of 0.1 and 0.13 for the ground and excited state, respectively, was used for the solid lines.

Figure 5 shows the calculated relative frequency of the four C transitions for SiV1 as a function of a magnetic field along the Z-axis, using the parameters given above and with equal orbital quenching factor of 0.1 for the ground and excited states (dotted lines) and with a quenching factor of 0.1 and 0.13 for the ground and excited state, respectively (solid lines). As can be seen from the figure, the difference in the quenching factors only causes small corresponding differences in the separations between relevant energy levels, since quenching factor matters only when magnetic field is closely aligned to the SiV axis. In this regard, the earlier studies are not expected to be sensitive to the relatively small differences in the quenching factors.

**5) Conclusion**

By introducing two coupling rates that can describe the combined effects of strain and JT coupling, we show that the energy level structure of a SiV center can be characterized with two relatively straightforward measurements under a transverse magnetic field, the CPT resonance position related to the two lower spin states in the ground state doublet and the frequency separation between two spin-conserved transitions in a PLE spectrum. The two strain/JT coupling rates can be derived from these measurements with the use of an analytical expression for the SiV energy level structure in a transverse magnetic field. Additional studies also reveal a slight difference in the orbital quenching factors between the ground and excited states, which can affect the energy level structure when the magnetic field is along or near the SiV axis. A full characterization of the SiV energy level structure is important for studies, such as acoustic quantum control, quantum spin-mechanics, or cavity QED, where an off-axis magnetic field needs to be used, and is also essential to the development of experiment approaches that can mitigate the variations of the SiV ground state energy levels due to the local strain environment.

**Acknowledgement**

The research effort at the University of Oregon has been supported by NSF under Grant Nos. 2012524 and 2003074. The effort at the Oak Ridge National Laboratory has been supported by the U. S. Department of Energy, Office of Science, Basic Energy Sciences, Materials Sciences and Engineering Division.